\documentclass[aps,pra,twocolumn,superscriptaddress]{revtex4-1}
\usepackage[latin9]{inputenc}
\usepackage{amsmath}
\usepackage{amssymb}
\usepackage{nicefrac}
\usepackage{graphicx}
\usepackage[usenames,dvipsnames]{color}

\makeatletter
 
 \@ifundefined{textcolor}{}
 {%
   \definecolor{BLACK}{gray}{0}
   \definecolor{WHITE}{gray}{1}
   \definecolor{RED}{rgb}{1,0,0}
   \definecolor{GREEN}{rgb}{0,1,0}
   \definecolor{BLUE}{rgb}{0,0,1}
   \definecolor{CYAN}{cmyk}{1,0,0,0}
   \definecolor{MAGENTA}{cmyk}{0,1,0,0}
   \definecolor{YELLOW}{cmyk}{0,0,1,0}
 }

\makeatother

\begin{document}

\title{Multipartite monogamous relations for Entanglement and Discord}

\author{Jonhy S. S. Ferreira}
\affiliation{Instituto de F\'isica, Universidade Federal de Mato Grosso, 78060-900, Cuiab\'a MT, Brazil}
\author{Dav\'i Filenga}
\affiliation{Faculdade de Ci\^encias, UNESP - Universidade Estadual Paulista, Bauru, SP, 17033-360, Brazil}
\author{Marcio F. Cornelio}
\affiliation{Instituto de F\'isica, Universidade Federal de Mato Grosso, 78060-900, Cuiab\'a MT, Brazil}
\author{Felipe F. Fanchini}
\affiliation{Faculdade de Ci\^encias, UNESP - Universidade Estadual Paulista, Bauru, SP, 17033-360, Brazil}

\begin{abstract}
The distribution of quantum correlations in multipartite systems play a significant role in several aspects of the quantum information theory. While it is well known that these quantum correlations can not be freely distributed, the way that it is shared in multipartite system is an open problem even for small set of qubits. Based on new 
{monogamy-like}
relations between entanglement and discord for $n$-partite systems, we show how these correlations are distributed in general, determining distinct equalities and inequalities to the quantum discord and the entanglement of formation for arbitrary multipartite pure states.
\end{abstract}
\maketitle

\section{Introduction}

Entanglement of formation (EF) \cite{Bennett1996b,Horodecki2009,Cornelio2010b}
and quantum discord (QD) \cite{Ollivier2001,Modi2012} are acknowledged
measures of quantum correlation for bipartite systems. They were derived
from very distinct concepts, however. The entanglement of formation was
first introduced by Bennett {\it et al.}\ \cite{Bennett1996b} and the goal
was to quantify the cost of building entangled states by local operations
and classical communication (LOCC). 
On the other hand, QD was first introduced by Olivier and Zurek \cite{Ollivier2001} 
with the aim of measuring the non-classicality of bipartite quantum states.
It determines the amount of locally inaccessible information and is given by the difference between two classically equivalent forms of the mutual information \cite{Ollivier2001}.

Despite their conceptual differences it is known that, as entanglement, QD can not be freely shared \cite{Modi2012}.
Indeed, the way that quantum correlations is distributed forbids it to be maximally correlated with two parts simultaneously since a monogamous relation arises.
The first to explore this aspect was Coffman, Kundu and Wootters \cite{CKW}, obtaining, about seventeen years ago,
the famous inequality for the monogamy of squared concurrence.
There, a simple expression concerning how entanglement is distributed among a system composed of three qubits was deduced, which raised an exciting and wide research field. 

The 
{monogamy of entanglement} has fundamental implications in several fields of quantum physics. For example, the lack of monogamy is considered a huge obstacle to the implementation of quantum cryptography, where unconditional security relies on the fact that the spy does not have the skills to correlate with the trusted parts \cite{Renes2006,Masanes2009}.
On the other hand, the lack of monogamy also provides information that may help to understand the mysterious behaviour of black holes \cite{Almheiri2013}, which appears when attempting to combine quantum mechanics with general relativity. Moreover, the monogamy of quantum correlation was essential for proving that asymptotic cloning is equivalent to state estimation \cite{Bae2006} and making quantum key distribution secure \cite{Scarani2005}.

{In this way, to elucidate the way that quantum correlation is distributed in multipartite systems is certainly important for information processing and communication technologies in multi-user scenarios.}
Nevertheless, in spite of the enormous effort of the scientific community to understand how quantum correlation is distributed in general multi-partite systems, that remains an important open problem even in the case of small number of parts and space dimensions. 
It is exactly at this direction that we develop our work, presenting
{new monogamy-like equalities for}
quantum discord and entanglement of formation in arbitrary multi-partite systems. As we show, extending the conservative relation between EF and QD \cite{Fanchini2011} for multipartite systems, a general rule for the way how quantum discord is distributed emerges. Also, the way that EF and QD are distributed is shown to be deeply related in general multipartite states. 
As EF is a way to quantify the quantum communication needed to build a bipartite state and QD is a way to quantify the amount of information inaccessible by local operations, our results relate these two concepts in a new form.
Indeed, in a multipartite system, the amount of quantum communication needed in each bipartition sums up equal to the sum of information trapped in nonlocal correlations as measured by quantum discord.
 
The paper is organized as follows. In section II we review some concepts and results necessary for this work. In section III we show two {monogamy-like} inequalities for fourpartite systems and two {monogamy-like} equalities for five-partite systems. In section IV, we extend the results to multipartite systems, presenting not only a new  {monogamy-like} law between EF and QD, but also a general equality elucidating how QD is distributed in multipartite systems. We conclude our work in section V.

\section{Review and background}

First of all, we need to define the concepts of entanglement and discord
we are going to use. Nowadays, there exist many different measures
of entanglement \cite{Horodecki2009} as well as different measures
of quantum correlation \cite{Modi2012}. In this work, we use the
EF as the measure of entanglement, first
defined by Bennett {\it et al}.\ in Ref.\ \cite{Bennett1996b} and the original
QD first defined by Olivier and Zurek in Ref.\ \cite{Ollivier2001}.
Indeed, both definitions are crucial for our work, since our main results come from the Koashi and Winter relation \cite{Koashi2004}, a notorious equation that connects EF and QD. 

First, we introduce the EF which is defined in the paradigm of local
operations with classical communication (LOCC). In this paradigm,
two spatially separated observers, usually called Alice and Bob, share
many copies of a standard quantum state. They can manipulate their
parts locally with arbitrary quantum operations and measurements and
communicate classically with each other. In this context, we consider
the problem of Alice and Bob having to build a particular mixed quantum
state $\rho_{ab}$ from the standard resource state, a maximally entangled
state of two qubits,
\[
	\left|\Phi\right\rangle =\frac{1}{\sqrt{2}}(\left|00\right\rangle +\left|11\right\rangle ).
\]
The first attempt to quantify this task arises with the entanglement
of formation \cite{Bennett1996b}. For any pure state
$\left|\psi_{ab}\right\rangle$,
it is easily evaluated through the von Neumann entropy of one of its
parts \cite{Bennett1996a,Nielsen2000}, 
\[
	E_{ab}(\left|\psi_{ab}\right\rangle )=S(\rho_{a}),
\]
where $\rho_{a}$ is the reduced state of subsystem $a$. For a mixed
state $\rho_{ab}$, EF is defined as
\[
	E_{ab}(\rho_{ab})=\min_{\mathcal{E}}\Big\{\sum_{i}p_{i}E_{ab}(\left|\varphi_{i}\right\rangle )\Big\},
\]
where the minimization runs over all possible ensembles such that
$\rho_{ab}=\sum_{i}p_{i}\left|\varphi_{i}\right\rangle \left\langle \varphi_{i}\right|$.
The EF can be easily calculated for two qubits states through Wootters's
formula \cite{Wooters1998}, however this calculation is very difficult for higher
dimensional system
\footnote{Today, it is known that the ultimate cost of building a state $\rho_{ab}$
from the standard state 
turns out to be the entanglement
cost which is the regularized entanglement of formation 
\cite{Bennett1996b,Horodecki2009,Hastings2009,Cornelio2010b}.
}.
Another way to think about the entangled is remembering that each standard state needs one bit of quantum communication to be formed. Therefore, EF is also a measure of the amount of quantum communication between Alice and Bob needed to build the state $\rho_{ab}$ \cite{Shor2004,Hastings2009,Cornelio2010b,Horodecki2009}. This interpretation will be useful in the following.

Second, we turn our attention to discord which aims to quantify the
amount of information that is not accessible in (or is destroyed by)
a measurement. Let us consider again a bipartite system with an arbitrary
state $\rho_{ac}$ shared between two observers, Alice and Carol.
The amount of uncertainty about subsystem $a$ is given by the von
Neumann entropy of this subsystem $S(\rho_{a})$. If Carol makes a
measurement on her subsystem $c$ and obtain a result $\Pi_{i}$ from
a complete set of POVM $\{\Pi_{i}\}$, the state of subsystem $a$
changes to the state $\rho_{a}^{i}$ with a new uncertainty given
by $S(\rho_{a}^{i})$. So the difference between $S(\rho_{a})$ and
$S(\rho_{a}^{i})$ is the amount of information learned by Carol about
the subsystem $a$. On average, Carol learns 
\[
S(\rho_{a})-\sum_{i}p_{i}S(\rho_{a}^{i}),
\]
where $p_{i}$ is the probability of Carol find the result $\Pi_i$
in her measurement from a complete set $\{\Pi_i\}$.
Of course, there are many possible sets of POVM Carol can choose
and she can always take the best one.
So the maximum amount of information she can obtain is called classical correlation
and is given by \cite{Henderson2001,Ollivier2001}
\begin{equation}	\label{eq:ClassicalCorrelation}
	J_{a|c}^{\leftarrow}(\rho_{ac})=\max_{\{\Pi_{i}\}}\big{[}S(\rho_{a})-\sum_{i}p_{i}S(\rho_{a}^{i})\big],
\end{equation}
where the maximization runs over all possible POVM on Carol's subsystem.
We remark that $J_{a|c}^{\leftarrow}$ is in general asymmetric being different
from $J_{c|a}^{\leftarrow}$. The classical correlation measure (\ref{eq:ClassicalCorrelation})
usually does not capture all the correlations contained in the quantum
state $\rho_{ac}$. The total amount of correlation is measured by
the quantum mutual information \cite{Nielsen2000}
\[
I(\rho_{ac})=S(\rho_{a})+S(\rho_{c})-S(\rho_{ac}),
\]
which is always bigger than $J_{a|c}(\rho_{ac})$. In this context,
the difference is the amount of inaccessible information which is measured
by the quantum discord \cite{Ollivier2001}
\[
\delta_{a|c}^{\leftarrow}(\rho_{ac})=I(\rho_{ac})-J_{a|c}^{\leftarrow}(\rho_{ac}).
\]
In addition, the amount of correlations $\delta_{a|c}^{\leftarrow}(\rho_{ac})$
is the amount that is destroyed when the measurement $\{\Pi_{i}\}$
is made in the Carol's subsystem. When the discord vanishes, no correlations
is destroyed by the measurement and the state does not change. In
this case, the state is considered classical \cite{Ollivier2001},
since it is a fundamental aspect of quantum mechanics to perturb the
physical systems in a measurement.

The entanglement of formation and the classical correlation are directly
connected in a pure tri-partite system by the Koashi-Winter (KW) relation
\cite{Koashi2004}, 
\begin{equation}
	E_{ab}+J_{a|c}^{\leftarrow}=S_{a},\label{eq:KoashiWinter}
\end{equation}
which express a kind of monogamy between these distinct measures (from now on, for brevity and clarity, we omit the quantum state between parenthesis, \emph{i.e.}\ $E_{ab}\equiv E_{ab}(\rho_{ab})$, $J_{a|c}^{\leftarrow} \equiv J_{a|c}^{\leftarrow}(\rho_{ac})$, etc.).
It means that the amount of quantum
correlations given by the EF between Alice and Bob, plus classical
correlation between Alice and Carol, is equal to the uncertainty about
the Alice's system. Also, Eq.\ (\ref{eq:KoashiWinter}) can easily be rewritten
to relate entanglement and discord as \cite{Fanchini2011}
\begin{equation}
	E_{ab}=\delta_{a|c}^{\leftarrow}+S_{a|c},
	\label{eq:fundamental}
\end{equation}
where $S_{a|c}$ is the quantum conditional entropy of the part $a$ given $c$, $S_{a|c} = S(\rho_{ac}) - S(\rho_c)$.
From cyclic permutations of Eq.\ (\ref{eq:fundamental}) one can obtain \cite{Fanchini2011}
\begin{equation}
	E_{ab}+E_{ac}=\delta_{a|b}^{\leftarrow}+\delta_{a|c}^{\leftarrow}.
	\label{eq:ConservationLaw}
\end{equation}
Therefore, the amount of entanglement a central particle $a$ shares
with the other two is equal to the amount of discord it also shares
with the same two particles. For this reason, it is called a conservation
law between entanglement and discord.
The entanglement and discord in a particular bipartition can be different,
nonetheless the amount of quantum correlation distributed through the entire system is the same measured either by entanglement or discord.

Moreover, one can also write another type of conservation law between entanglement and discord
involving a kind of cycle over the parts of the system \cite{Fanchini2012}
\begin{equation}
	E_{ab} + E_{bc} + E_{ca} = \delta_{b|a}^{\leftarrow} + \delta_{c|b}^{\leftarrow} + \delta_{a|c}^{\leftarrow} .
	\label{eq:LLI}
\end{equation}
The cycle appears in the measurements involving the definition of discord.
We have a measurement in part $a$ referring to $b$,
then one in $b$ to $c$ following by one in $c$ to $a$ closing the cycle.
{Although Eq.\ (\ref{eq:LLI}) is already known, we are going to look at it with a new point of view, which is particularly more interesting for the generalizations gotten in this work. The right side is the sum of EFs, so it is the sum of quantum communications need to simulated the correlation present in each of one of bipartitions. The left side is the sum of discords in each bipartition, so the left side is the sum of in inaccessible information of each bipartition. Therefore, the sum of quantum communication needed in each bipartition is equal to the sum of information that is trapped in non-local correlations.
}

Moreover, Eq.\ (\ref{eq:LLI}) can be written in the opposite direction,
since it is not different from a permutation of the parts.
Therefore, as the entanglements are symmetric quantities,
we can write a conservation law for discords only \cite{Fanchini2012}
\begin{equation}
 	\delta_{a|b}^{\leftarrow} + \delta_{b|c}^{\leftarrow} + \delta_{c|a}^{\leftarrow}
 	= \delta_{b|a}^{\leftarrow} + \delta_{c|b}^{\leftarrow} + \delta_{a|c}^{\leftarrow}.
	\label{eq:LLI_discord}
\end{equation}
Eq.\ (\ref{eq:LLI_discord}) shows that the sum of locally inaccessible information is the same in the two possible directions of the cycle.
The Eqs.\ (\ref{eq:LLI}) and (\ref{eq:LLI_discord}) are generalized to multipartite systems in section \ref{generalresults}.

\section{Entanglement and discord in four and five-partite systems}		\label{ParticularResults}

Given this brief overview, we are now in position to discuss the generalization of our previous results \cite{Fanchini2011,Fanchini2012}. We start with four-partite system, however straight forward generalization  gives inequalities instead of equalities found in three-partite systems,  {Eqs.\ (\ref{eq:ConservationLaw}), (\ref{eq:LLI}) and (\ref{eq:LLI_discord})}. These inequalities relates entanglement, discord and the strong subadditivity inequality which is interesting also, but they do not result in new conservation laws. Nevertheless, generalization to five-partite systems do result in equalities and new conservation laws. These results shed light on how to generalized conservations laws for arbitrary dimensions. As we discuss in the following, some difficult arises because there is a difference between even and odd number of parts. The generalizations are discussed in section \ref{generalresults}.


\subsection{Four-partite systems}

\subsubsection{Inequalities with a central particle}

From three-partite systems to four-partite, one could expect to
find equalities between sums of EF and QD.
Nevertheless, what we found are inequalities which are closed related to the strong subadditivity of von Neumann entropy.
The difficulty arises in dividing properly the parts of the system, since there is more ways to do it in four than in three partite systems.
A first attempt, from Eq.\ (\ref{eq:fundamental}),
is to write the following equations,
\begin{eqnarray}
	E_{a|bc} & = & \delta_{a|d}^{\leftarrow}+S_{a|d},\nonumber \\
	E_{a|cd} & = & \delta_{a|b}^{\leftarrow}+S_{a|b}.
	\label{eq:fundfourparts}
\end{eqnarray}
Combining Eqs.\ (\ref{eq:fundfourparts}), we get
\begin{equation}
	E_{a|bc}+E_{a|cd}=\delta_{a|d}^{\leftarrow}+\delta_{a|b}^{\leftarrow}+S_{a|d}+S_{a|b}.
	\label{eq:sum4parts}
\end{equation}
The sum of conditional entropies can be rewritten as a strong subadditivity inequality,
\[
	S_{a|d}+S_{a|b}=S_{ab}-S_{b}+S_{bc}-S_{abc} \geq 0.
\]
Therefore, the sum of conditional entropies does not cancel.
Nevertheless, we get the following inequality for the distribution of entanglement and discord in four partite systems
\begin{equation}
	E_{a|bc}+E_{a|cd} \geq \delta_{a|d}^{\leftarrow}+\delta_{a|b}^{\leftarrow}.
	\label{eq:fourpartsEgeD}
\end{equation}

Moreover, instead of Eq.\ (\ref{eq:fundfourparts}), we can write the fundamental relations in the following form
\begin{eqnarray*}
E_{ab} & = & \delta_{a|cd}^{\leftarrow}+S_{a|cd},\\
E_{ad} & = & \delta_{a|bc}^{\leftarrow}+S_{a|bc}.
\end{eqnarray*}
Therefore, with a similar reasoning, we obtain the complementary inequality
\begin{equation}
	E_{ab}+E_{ad} \leq \delta_{a|bc}^{\leftarrow} + \delta_{a|cd}^{\leftarrow}.
	\label{eq:fourpartsEleD}
\end{equation}
In Eqs.\ (\ref{eq:fourpartsEgeD}) and (\ref{eq:fourpartsEleD}), we
notice that particle $a$ is present in all quantities being the central
one of the relation. We also notice the change in particle $c$.
In Eq.\ (\ref{eq:fourpartsEgeD}), it is in the entanglements while,
in Eq.\ (\ref{eq:fourpartsEleD}), it is in the discords.
{That is, part $c$ is always in the greater side of inequality which is, in fact, reasonable,
once that tracing out a quantum system always decreases entanglement and, most of times, also discord.}

In addition, it is possible to obtain inequalities involving all possible
entanglements with the part $a$.
This can be done starting from the following fundamental relations
\begin{eqnarray}
	E_{a|bc} & = & \delta_{a|d}+S_{a|d},\nonumber \\
	E_{a|bd} & = & \delta_{a|c}+S_{a|c},\nonumber \\
	E_{a|cd} & = & \delta_{a|b}+S_{a|b}.
	\label{eq:fourparts3terms}
\end{eqnarray}
Combining Eqs.\ (\ref{eq:fourparts3terms}), we obtain
\begin{multline}
	E_{a|bc}+E_{a|cd}+E_{a|db}\\
	=\delta_{a|b}^{\leftarrow} +\delta_{a|c}^{\leftarrow} +\delta_{a|d}^{\leftarrow} +S_{a|b} +S_{a|c} +S_{a|d}.
	\label{eq:Sum3terms}
\end{multline}
The sum of conditional entropies is always positive, since it is a combination of three strong subadditivity inequalities \cite{Nielsen2000}. Therefore we have 
\begin{equation}
	E_{a|bc}+E_{a|cd}+E_{a|db}\geq\delta_{a|b}^{\leftarrow}+\delta_{a|c}^{\leftarrow}+\delta_{a|d}^{\leftarrow}.
	\label{12}
\end{equation}

With a similar reasoning to the Eq.\ (\ref{eq:fourpartsEleD}), we also have the complementary inequality
\begin{equation}
	E_{ab}+E_{ac}+E_{ad}\leq\delta_{a|bc}^{\leftarrow}+\delta_{a|cd}^{\leftarrow}+\delta_{a|cd}^{\leftarrow}.
	\label{menor}
\end{equation}

These results show how distinct inequalities between EF and QD emerge when treating one of the particles as a central one, similarly to Eq.\ (\ref{eq:ConservationLaw}). In the following, we treat similar aspects, but now cycling the parts in discord side, similarly to Eq.\ (\ref{eq:LLI}).

\subsubsection{Cycling inequalities}

All the relations (\ref{eq:fourpartsEgeD}, \ref{eq:fourpartsEleD}, \ref{12}, \ref{menor}) involves a central particle
that is present in all quantities (particle $a$). So These inequalities are related to the generalization of the Eq.\ (\ref{eq:ConservationLaw}).
Here we derive another type of inequalities which involve cycling the parts in discord side and is related to the generalization of the Eq.\ (\ref{eq:LLI}).
This forms a cycle of inaccessible local information as is discussed in Ref.\ \cite{Fanchini2012},
but now with four parts. First we consider the cycle $a\rightarrow b$,
$b\rightarrow c$, $c\rightarrow d$ and $d\rightarrow a$ of local
inaccessible information or discords, writing down the following
fundamental equations (\ref{eq:fundamental}),
\begin{eqnarray}
	E_{b|cd} & = & \delta_{b|a}^{\leftarrow}+S_{b|a},\nonumber \\
	E_{c|da} & = & \delta_{c|b}^{\leftarrow}+S_{c|b},\nonumber \\
	E_{d|ab} & = & \delta_{d|c}^{\leftarrow}+S_{d|c},\nonumber \\
	E_{a|bc} & = & \delta_{a|d}^{\leftarrow}+S_{a|d}.
	\label{eq:4partsFundamentals}
\end{eqnarray}
Combining Eqs.\ (\ref{eq:4partsFundamentals}), we get
\begin{multline*}
	E_{a|bc}+E_{b|cd}+E_{c|da}+E_{d|ab}\\
	=\delta_{a|d}^{\leftarrow} +\delta_{d|c}^{\leftarrow} +\delta_{c|b}^{\leftarrow} +\delta_{b|a}^{\leftarrow} +S_{a|d}+S_{b|a}+S_{c|b}+S_{d|c}.
\end{multline*}
The sum of conditionals entropies is always positive, due to the strong
subadditivity inequality. This can be seen after some manipulation
of the entropies,
\begin{multline*}
	S_{a|d}+S_{b|a}+S_{c|b}+S_{d|c}\\
	=(S_{ab}+S_{bc}-S_{b}-S_{abd})+(S_{bc}+S_{cd}-S_{bcd}-S_{c}) \geq 0.
\end{multline*}
Therefore, we are left with the following relation between entanglements
and discords,
\begin{equation}
	E_{a|bc}+E_{b|cd}+E_{c|da} +E_{d|ab}\geq\delta_{a|d}^{\leftarrow} +\delta_{d|c}^{\leftarrow}+\delta_{c|b}^{\leftarrow} +\delta_{b|a}^{\leftarrow}.
	\label{4ciclomaior}
\end{equation}

We can also construct a cycle of LII using a different setup of parts.
Let the cycle $ab\rightarrow c,$ $cb\rightarrow d$, $cd\rightarrow a$
and $da\rightarrow b$. This allows us to write the following relations
between entanglements and discords from Eq. (\ref{eq:fundamental}),
\begin{eqnarray}
	E_{cd} & = & \delta_{c|ab}^{\leftarrow}+S_{c|ab},\nonumber \\
	E_{da} & = & \delta_{d|bc}^{\leftarrow}+S_{d|bc},\nonumber \\
	E_{ab} & = & \delta_{a|cd}^{\leftarrow}+S_{a|cd},\nonumber \\
	E_{bc} & = & \delta_{b|ad}^{\leftarrow}+S_{b|ad}.
	\label{eq:4partsCycle2}
\end{eqnarray}
With a similar manipulation, the sum of conditional entropies in Eqs.\
(\ref{eq:4partsCycle2}) can be shown to be always positive due to
the strong subadditivity inequality. Therefore, we are left with
the following inequality between entanglements and discords for pure
four-partite systems
\begin{equation}
	E_{ab} +E_{bc} +E_{cd} +E_{da} \leq \delta_{a|cd}^{\leftarrow} +\delta_{b|da}^{\leftarrow} +\delta_{c|ab}^{\leftarrow} +\delta_{d|bc}^{\leftarrow}.
	\label{4ciclomenor}
\end{equation}

Eqs.\ (\ref{eq:4partsCycle2}) and (\ref{4ciclomenor}) above finalize our results considering fourpartite systems and cycling relations.

%

\subsection{Five-partite systems}

For pure five-partite systems, we return to find equalities instead
of inequalities in a very similarly to what happens in the three-partite
case,

\subsubsection{Equalities with a central particle}

From conservation law Eq.\  (\ref{eq:ConservationLaw}) we can write
the following three equalities for five partite systems
\begin{eqnarray}
	E_{a|bc}+E_{a|de} & = & \delta_{a|bc}^{\leftarrow}+\delta_{a|de}^{\leftarrow},\nonumber \\
	E_{a|bd}+E_{a|ce} & = & \delta_{a|bd}^{\leftarrow}+\delta_{a|ce}^{\leftarrow},\nonumber \\
	E_{a|be}+E_{a|cd} & = & \delta_{a|be}^{\leftarrow}+\delta_{a|cd}^{\leftarrow}.
	\label{eq:5partsConservationsLaws}
\end{eqnarray}
So, we can write down the equality
\begin{multline}
	E_{a|bc}+E_{a|de}+E_{a|bd}+E_{a|ce}+E_{a|be}+E_{a|cd}\\
	=\delta_{a|de}^{\leftarrow}+\delta_{a|bc}^{\leftarrow}+\delta_{a|ce}^{\leftarrow}+\delta_{a|bd}^{\leftarrow}+\delta_{a|cd}^{\leftarrow}+\delta_{a|be}^{\leftarrow}
	\label{eq:ConservationLaw5parts}
\end{multline}
In Eq.\ (\ref{eq:ConservationLaw5parts}) one can check that we have
all combinations of entanglements and discords of the particle $a$
with all other possible combinations of two particles.
Therefore, similarly what happens in three-partite
case, Eq.\ (\ref{eq:ConservationLaw}), the sum of entanglements a central particle, $a$,
shares with all the others possible combinations of two particles
is equal to the sum of all discords between the same bipartitions,
establishing a {monogamy-like} conservation law of quantum correlations.
 
Despite this fact, it is also evident that Eq.\ (\ref{eq:ConservationLaw5parts}) is a weaker statement than the three Eqs.\ (\ref{eq:5partsConservationsLaws}). 
In this way, although it is possible to derive generalizations like Eq.\ (\ref{eq:ConservationLaw5parts}), they follow from straight forward combinations of the three partite conservation law Eq.\ (\ref{eq:ConservationLaw}) with the appropriate combination of subsystems, Eqs.\ (\ref{eq:5partsConservationsLaws}). 
Nevertheless, as we show bellow, nontrivial results emerges when considering cycling equalities in five-partite systems.

\subsubsection{A cycling equality}		\label{5partscycle}

More interesting are the equality that arises when we generalize Eq.\ (\ref{eq:LLI}) for five-partite systems, 
since they are not direct application of the similar conservation law from system of small number of parts, but really new conservation laws.
From the fundamental Eq.\ (\ref{eq:fundamental}), we can write the
following equations,
\begin{eqnarray}		\label{eqs:fund5patrs}
	E_{a|bc} & = & \delta_{a|ed}^{\leftarrow}+S_{aed}-S_{ed},\\
	E_{c|de} & = & \delta_{c|ba}^{\leftarrow}+S_{cba}-S_{ba},\\
	E_{e|ab} & = & \delta_{e|dc}^{\leftarrow}+S_{edc}-S_{dc}.\\
	E_{b|cd} & = & \delta_{b|ae}^{\leftarrow}+S_{bae}-S_{ae},\\
	E_{d|ea} & = & \delta_{d|cb}^{\leftarrow}+S_{dcd}-S_{cb},
\end{eqnarray}
Notice that the entropy $S_{de}$ from the first equation cancels
with $S_{cba}$ from the second one; the entropy $S_{ba}$ from the second equation cancels with $S_{ecd}$ from third one and so on.
When we sum all of the equations, the cycle closes and all the entropies cancel out. The result is the following equality
\begin{multline}
	E_{a|bc}+E_{b|cd}+E_{c|de}+E_{d|ea}+E_{e|ab}\\
	=\delta_{a|de}^{\leftarrow}+\delta_{b|ea}^{\leftarrow}+\delta_{c|ab}^{\leftarrow}+\delta_{d|bc}^{\leftarrow}+\delta_{e|cd}^{\leftarrow}.
	\label{eq:5partscycle}
\end{multline}
Eq.\ (\ref{eq:5partscycle}) contains a cycle of LII information
in the right side similarly what happens in the three-partite case Eq.\ (\ref{eq:LLI}). As we show in section IV, these results gives the direction for a generalization to the multipartite case.

\section{Generalized Conservation Laws for multipartite systems}		\label{generalresults}

Now, with the particular results presented in the section \ref{ParticularResults}, we extend our results for multipartite systems.

\begin{widetext}
\subsection{Generalized cycling conservation laws}

The idea behind the deduction of Eq.\ (\ref{eq:5partscycle}) is the key idea for generalizing it to multipartite systems. Let us consider a system composed by $N$ parts, where $N$ is odd, and let $n =  \nicefrac{(N-1)}{2}$. So we can write down the following equations
\begin{equation}
\begin{array}{rclllll}		\label{eqs:fund_for_conservaionalaw_odd}
	E_{1|2,3,\dots,n+1} & = &\delta_{1|N,N-1,\dots,n+2}^\leftarrow &+& S_{1,N,N-1,\dots,n+2} &-& S_{N,N-1,\dots,n+2}, \\
	E_{n+1|n+2,n+3,\dots,N} & = &\delta_{n+1|n,n-1,\dots,1}^\leftarrow &+& S_{n+1,n,\dots,1} &-& S_{n,n-1,\dots,1}, \\
	E_{N|1,2,\dots,n} & = &\delta_{N|N-1,N-2,\dots,n+1}^\leftarrow &+& S_{N,N-1,\dots,n+1} &-& S_{N-1,N-2,\dots,n+1}, \\
	E_{n|n+1,n+2,\dots,N-1} & = &\delta_{n|n-1,n-2,\dots,N}^\leftarrow &+& S_{n,n-1,\dots,N} &-& S_{n-1,n-2,\dots,1,N}, \\
	E_{N-1|N,1,\dots,n-1} & = &\delta_{N-1|N-2,N-3,\dots,n}^\leftarrow &+& S_{N-1,N-2,\dots,n} &-& S_{N-2,N-3,\dots,n}, \\
	\vdots \hspace{1cm} & = &  \hspace{1cm} \vdots  \\
	E_{n+2|n+3,n+4,\dots,1} & = &\delta_{n+2|n+1,n,\dots,2}^\leftarrow &+& S_{n+2,n+1,n,\dots,2} &-& S_{n+1,n,\dots,2}, 
\end{array}
\end{equation}
In the Eqs. (\ref{eqs:fund_for_conservaionalaw_odd}), the next equations is always based on $n$ subsystems at right from the previous one. We can check that, when the equations are summed up, the second entropy cancels with the first one from next equation until the cycle is closed. The result is the following conservation law
\begin{multline}	\label{eq:GeneralConservationLawOdd}
	E_{1|2,3,\dots,n+1} + E_{2|3,4,\dots,n+2} + \cdots + E_{N|1,2,\dots,n}
	= \delta_{1|N,N-1,\dots,n+2}^\leftarrow + \delta_{2|1,N,N-1,\dots,n+3}^\leftarrow
	 + \cdots + \delta_{N|N-1,N-2,\dots,n+1}^\leftarrow.
\end{multline}
{Eq.\ (\ref{eq:GeneralConservationLawOdd}) is the generalization of Eq.\ (\ref{eq:LLI}). It shows how the amount of quantum communication needed in each bipartition sums up equal to the sum of information trapped in nonlocal correlations as measured by quantum discord. In this way, in order of obtaining an equality for these two quantities, Eq.\ (\ref{eq:GeneralConservationLawOdd}) shows that we must organize the sum of entanglements forming a cycle including all the bipartition in one direction and the sum of discords in the opposite direction, when the number of subsystems is odd.
}


At first sight, it appears that the idea behind derivation of the conservation law (\ref{eq:GeneralConservationLawOdd}) only works for systems with an odd number of parts, since only in these cases the entropies have the right numbers of parts to cancel each other.
Nevertheless, it is still possible to make them cancel each other varying the number of subsystems.
Let us come back to the four-partite case, for instance.
Then we write down the following sequence of Equations, alternating the number of parts in entanglements and discords,
\begin{eqnarray*}
	E_{a|bc} & = & \delta_{a|d}^{\leftarrow}+S_{ad}-S_{d},\\
	E_{c|d}	 & = & \delta_{c|ba}^{\leftarrow}+S_{cba}-S_{ba},\\
	E_{d|ab} & = & \delta_{d|c}^{\leftarrow}+S_{dc}-S_{c},\\
	E_{b|c}  & = & \delta_{b|ad}^{\leftarrow}+S_{bad}-S_{ad},
\end{eqnarray*}
One can check that again the second entropy cancels with the first entropy of the next equations, and the cycle closes at the last one. Summing up these equations results
\begin{equation}		\label{eq:ConservationLaw4parts}
	E_{a|bc} + 	E_{c|d}	+ E_{d|ab} + E_{b|c} =
	\delta_{a|d}^{\leftarrow} + \delta_{c|ba}^{\leftarrow}+ \delta_{d|c}^{\leftarrow}+ \delta_{b|ad}^{\leftarrow}.
\end{equation}
This equation is a conservation law for entanglement and discord for four-partite systems where the number of parts changes. Therefore we see that it is possible to derive conservation laws for multipartite systems composed of an even number of parts, but the number of parts in the bipartitions must vary accordingly to Eq.\ (\ref{eq:ConservationLaw4parts}).

Nonetheless, the Eq. (\ref{eq:ConservationLaw4parts}) does not show the full general rule for deriving a general conservation law for even number of parts. This appears when we write down the equations for a six-partite system,
\begin{align}		\label{eqs:fund_for_conservation_law_6parts}
	E_{1|23}  &= \delta_{1|654}^\leftarrow + S_{1654} - S_{654}, & E_{4|56} &= \delta_{4|321}^\leftarrow + S_{4321} - S_{321}, \nonumber \\
	E_{3|456} &= \delta_{3|21}^\leftarrow + S_{321} - S_{21}, & E_{6|123} &= \delta_{6|54}^\leftarrow + S_{654} - S_{54},	 \nonumber \\
	E_{6|12} &= \delta_{6|543}^\leftarrow + S_{6543} - S_{543}, & E_{3|45} &= \delta_{3|216}^\leftarrow + S_{3216} - S_{216}, \nonumber \\
	E_{2|345} &= \delta_{2|16}^\leftarrow + S_{216} - S_{16}, & E_{5|612} &= \delta_{5|43}^\leftarrow + S_{543} - S_{43},	 \nonumber	\\
	E_{5|61} &= \delta_{5|432}^\leftarrow + S_{5432} - S_{432}, & E_{2|34} &= \delta_{2|165}^\leftarrow + S_{2165} - S_{165}, \nonumber \\
	E_{1|234} &= \delta_{1|65}^\leftarrow + S_{165} - S_{65},& E_{4|561} &= \delta_{4|32}^\leftarrow + S_{432} - S_{32}.
\end{align}
Similarly with happens in the previous cases, when all the Eqs.\ (\ref{eqs:fund_for_conservation_law_6parts})  are summed up, the negative entropies cancel with the positive one from the next equation. The cycle closes when the negative entropy from the last equation cancels with the positive one from the first. The number of parts in the bigger bipartitions alternates between $n$ and $n-1$. The result is the following conservation law:
\begin{multline}		\label{eq:Conservaition_law_6parts}
	E_{1|23} + E_{1|234} + E_{2|34} + E_{2|345} + E_{3|45} + E_{3|456} + E_{4|56} + E_{4|561} + E_{5|61} + E_{5|612} + E_{6|12} + E_{6|123}\\
	= \delta_{1|654}^\leftarrow + \delta_{1|65}^\leftarrow + \delta_{2|165}^\leftarrow + \delta_{2|16}^\leftarrow + \delta_{3|216}^\leftarrow + \delta_{3|21}^\leftarrow + \delta_{4|321}^\leftarrow + \delta_{4|32}^\leftarrow + \delta_{5|432}^\leftarrow +\delta_{5|43}^\leftarrow + \delta_{6|543}^\leftarrow + \delta_{6|543}^\leftarrow
\end{multline}

From Eqs.\ (\ref{eqs:fund_for_conservation_law_6parts}) and Eq.\ (\ref{eq:Conservaition_law_6parts}) it is easy to state the general conservation law for multipartite systems with an even number of parts, $N$,
\begin{multline}		\label{eq:Conservation_law_even}
	E_{1|23\dots n} + E_{1|23\dots,n+1} + E_{2|34\dots n+1} + E_{2|34 \dots n+2}
 + \cdots + E_{N|12\dots n-1} + E_{N|12\dots n} \\
	= \delta_{1|N-1,N-2,\dots,n+1}^\leftarrow + \delta_{1|N-1,N-2,\dots,n}^\leftarrow
	+ \delta_{2|1,N-1,N-2,\dots,n+2}^\leftarrow + \delta_{2|1,N-1,N-2,\dots,n+1}^\leftarrow\\
	+ \cdots + \delta_{N|N-1,N-2,\dots,n}^\leftarrow + \delta_{N|N-1,N-2,\dots,n-1}^\leftarrow 
\end{multline}
where $n=\nicefrac{N}{2}$. The Eq.\ (\ref{eq:GeneralConservationLawOdd}) and Eq.\ (\ref{eq:Conservation_law_even}) are the natural generalization of Eq.\ (\ref{eq:LLI}) for an even number of parts. 
{ The organization of the sum of terms in Eq.\ (\ref{eq:Conservation_law_even}) is more involving than in Eq.\ (\ref{eq:GeneralConservationLawOdd}), however the same interpretation applies. 
On the left side we have a sum of entanglements which represent the amount of quantum communication needed to form the correlation in each bipartition while, in the right side, the discords represent the amount of information that is not accessible locally.}	
These two equations show, for general multipartite pure states, how EF and QD are distributed in a simpler and intuitive expression, 
{associating the amount of quantum communication needed to form the correlations in bipartitions with the amount of correlations which is inaccessible by local operations.}

\end{widetext}

\subsection{Generalized Conservation Law for Discord}

In this section, we now focus on how QD is distributed, extending Eq.\ (\ref{eq:LLI_discord}) for multipartite systems. Here a closed expression, based purely on QD, is deduced demonstrating the way that quantum correlations is distributed for general pure states.
To determine a closed form to the discord distribution in multipartite systems, we begin considering a four-partite system. Firstly, we note that, by means of the KW relation, we can write
\begin{eqnarray}
	\delta_{b|a}^{\leftarrow}  & = & E_{b|cd} - S_{b|a},\nonumber \\
	\delta_{c|b}^{\leftarrow}  & = & E_{c|da} - S_{c|b},\nonumber \\
	\delta_{d|c}^{\leftarrow}  & = & E_{d|ab} - S_{d|c},\nonumber \\
	\delta_{a|d}^{\leftarrow}  & = & E_{a|bc} - S_{a|d},
	\label{4partsFundamentals}
\end{eqnarray}
since that combining the Eqs.\ (\ref{4partsFundamentals}), we get
\begin{multline}
	E_{a|bc}+E_{b|cd}+E_{c|da}+E_{d|ab}\\
	=\delta_{a|d}^{\leftarrow}+\delta_{d|c}^{\leftarrow}+\delta_{c|b}^{\leftarrow}+\delta_{b|a}^{\leftarrow}+S_{a|d}+S_{b|a}+S_{c|b}+S_{d|c}.
	\label{4partsD}
\end{multline}

The interesting aspect about the set of equations given by Eq.\ (\ref{4partsFundamentals}) is that it is possible to organize them, with Eq.\ (\ref{eq:4partsCycle2}), to obtain an equality between entanglement and discord even for 4-partite systems. Indeed, subtracting one to the other we see that 
\begin{multline}
	(E_{a|bc}\!+\!E_{ab}) + (E_{b|cd}\! +\!E_{bc})\\
 	+(E_{c|da}\! +\!E_{cd}) +(E_{d|ab}\!+\!E_{da})\\
 	= (\delta_{a|cd}^{\leftarrow}\!+\!\delta_{a|d}^{\leftarrow})  + (\delta_{b|ad}^{\leftarrow}\!+\! \delta_{b|a}^{\leftarrow}) \\ +(\delta_{c|ab}^{\leftarrow}\!+\!\delta_{c|b}^{\leftarrow}) + (\delta_{d|bc}^{\leftarrow}\!+\! \delta_{d|c}^{\leftarrow}).
	\label{law41}
\end{multline}
Eq.\ (\ref{law41}) is a conservative equation between EF and QD for 4-partite systems and it shows two curious aspects: firstly, we note that bipartitions of different sizes must be considered, since that three and two parts terms appears in Eq.\ (\ref{law41}). Secondly, we note that, contrary to the conservative relation for a tripartite or 5-partite pure state, there is no symmetry between left and right sides of the equation above. It is certainly a strange aspect which induce us to search for another conservative relation, a symmetric one. 

For this purpose, we consider a different set of equations.
Again we focus on the entanglement between two parts, but contrarily to the set of equations given in Eq.\ (\ref{eq:4partsCycle2}), we write the KW relation for $E_{ba}$, $E_{cb}$, $E_{dc}$, and $E_{ad}$.
It is clear, since entanglement is a symmetric entity, that these amounts of entanglement are equivalent to that given in Eq.\ (\ref{eq:4partsCycle2}), but with this we can derive a different set of equations:
\begin{eqnarray}
E_{ba} & = & \delta_{b|cd}^{\leftarrow}+S_{b|cd},\nonumber\\
E_{cb} & = & \delta_{c|ad}^{\leftarrow}+S_{c|ad},\nonumber\\
E_{dc} & = & \delta_{d|ab}^{\leftarrow}+S_{d|ab},\nonumber\\
E_{ad} & = & \delta_{a|bc}^{\leftarrow}+S_{a|bc}.
\label{4partsE2}
\end{eqnarray}
Again, combining the set of Eqs.\ (\ref{4partsE2}) above with that given in Eqs.\ (\ref{4partsFundamentals}), we derive another conservative relation:
\begin{multline}
	(E_{a|bc}\!+\!E_{ab}) + (E_{b|cd}\! +\!E_{bc})\\
	 + (E_{c|da}\! +\!E_{cd}) +(E_{d|ab}\!+\!E_{da})\\
	 = (\delta_{a|bc}^{\leftarrow}\!+\!\delta_{b|a}^{\leftarrow})  + (\delta_{b|cd}^{\leftarrow}\!+\! \delta_{c|b}^{\leftarrow})\\ +(\delta_{c|da}^{\leftarrow}\!+\!\delta_{d|c}^{\leftarrow}) + (\delta_{d|ab}^{\leftarrow}\!+\! \delta_{a|d}^{\leftarrow})
	\label{law42}
\end{multline}


Contrarily to the conservative equation given by Eq.\ (\ref{law41}), Eq.\ (\ref{law42}) above is symmetric and, surprisingly, when combined we obtain a rule for the distribution of QD for a 4-partite system:
\begin{multline}
	\delta_{a|bc}^{\leftarrow}+\delta_{b|cd}^{\leftarrow}+\delta_{c|ad}^{\leftarrow}+\delta_{d|ab}^{\leftarrow} \\
	 = \delta_{a|cd}^{\leftarrow}+\delta_{b|ad}^{\leftarrow}+\delta_{c|ab}^{\leftarrow}+\delta_{d|bc}^{\leftarrow}.
	\label{mono4}
\end{multline}
This equation elucidates an interesting property about the way that quantum correlations is shared in a four-partite systems. It shows that the total inaccessible information, about individual parts, after bipartite cyclical measurements (\textit{i.e.}\ $\delta_{a|bc}^{\leftarrow}+\delta_{b|cd}^{\leftarrow}+\delta_{c|ad}^{\leftarrow}+\delta_{d|ab}^{\leftarrow}$), is equivalent to the total inaccessible information after counter-cyclic measurements. Also, it is important to note that Eq.\ (\ref{mono4}) could be directly obtained rearranging the set of equation given in Eq.\ (\ref{eq:4partsCycle2}) and Eq.\ (\ref{4partsE2}), but we choose this manner, since the conservative equation between EOF and QD is also deduced.

Now, continuing our endeavor to generalize the result above for multipartite systems, we explore the 5-partite systems. As usual, we begin with three set of equations:
\begin{eqnarray}
E_{a|bcd} & = & \delta_{a|e}^{\leftarrow}+S_{a|e},\nonumber\\
E_{b|cde} & = & \delta_{b|a}^{\leftarrow}+S_{b|a},\nonumber\\
E_{c|dea} & = & \delta_{c|b}^{\leftarrow}+S_{c|b},\nonumber\\
E_{d|eab} & = & \delta_{d|c}^{\leftarrow}+S_{d|c},\nonumber\\
E_{e|abc} & = & \delta_{e|d}^{\leftarrow}+S_{e|d},
\label{five1}
\end{eqnarray}
\begin{eqnarray}
E_{ab} & = & \delta_{a|cde}^{\leftarrow}+S_{a|cde},\nonumber\\
E_{bc} & = & \delta_{b|ade}^{\leftarrow}+S_{b|ade},\nonumber\\
E_{cd} & = & \delta_{c|abe}^{\leftarrow}+S_{c|abe},\nonumber\\
E_{de} & = & \delta_{d|abc}^{\leftarrow}+S_{d|abc},\nonumber\\
E_{ea} & = & \delta_{e|bcd}^{\leftarrow}+S_{e|bcd},
\label{five2}
\end{eqnarray}
and
\begin{eqnarray}
E_{ba} & = & \delta_{b|cde}^{\leftarrow}+S_{b|cde},\nonumber\\
E_{cb} & = & \delta_{c|ade}^{\leftarrow}+S_{c|ade},\nonumber\\
E_{dc} & = & \delta_{d|abe}^{\leftarrow}+S_{d|abe},\nonumber\\
E_{ed} & = & \delta_{e|abc}^{\leftarrow}+S_{e|abc},\nonumber\\
E_{ae} & = & \delta_{a|bcd}^{\leftarrow}+S_{a|bcd}.
\label{five3}
\end{eqnarray}
Combining the set of equations given by Eqs.\ ({\ref{five1}}) with the set of equations given by Eqs.\ (\ref{five3}), we directly obtain the conservative relation: 
\begin{multline}
	(E_{a|bcd}+E_{ae})+(E_{b|cde}+E_{ba})+(E_{c|dea}+E_{cb})\\
	+(E_{d|eab}+E_{dc})+(E_{e|abc}+E_{ed}) \\
	= (\delta_{a|bcd}^{\leftarrow}+\delta_{a|e}^{\leftarrow}) + (\delta_{b|cde}^{\leftarrow}+\delta_{b|a}^{\leftarrow}) + (\delta_{c|dea}^{\leftarrow}+\delta_{c|b}^{\leftarrow})\\
	 + (\delta_{d|eab}^{\leftarrow}+\delta_{d|c}^{\leftarrow}) + (\delta_{e|abc}^{\leftarrow}+\delta_{e|d}^{\leftarrow}).
\end{multline}
Also, combining the set of equation given by Eqs.\ (\ref{five2}) and Eqs.\ (\ref{five3}) we obtain the conservation law for discord
  for 5-partite systems: 
\begin{multline}
\delta_{a|cde}^{\leftarrow}+\delta_{b|ade}^{\leftarrow}+\delta_{c|abe}^{\leftarrow}+\delta_{d|abc}^{\leftarrow}+\delta_{e|bcd}^{\leftarrow} \\
= \delta_{a|bcd}^{\leftarrow}+\delta_{b|cde}^{\leftarrow}+\delta_{c|ade}^{\leftarrow}+\delta_{d|abe}^{\leftarrow}+\delta_{e|abc}^{\leftarrow}.
\label{mono5}
\end{multline}

\begin{figure} 
	\begin{center}
		\includegraphics{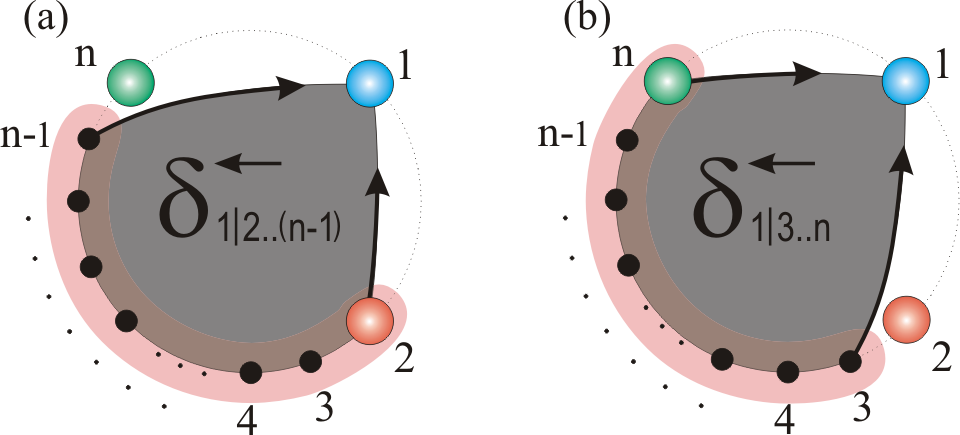} {}
	\end{center}
	\caption{(Color Online) (a) the inaccessible information of part $1$, given an observation on the rest, with exception of that immediately on its left, \emph{i.e.}\ part $n$. (b) the inaccessible information on part $1$, but now given an observation on the rest,  with exception of that immediately on its right, \emph{i.e.}\ part $2$. The monogamous equality is reached doing the same procedure for all $n$ parts.} 
	\label{fig1}
\end{figure}

With the results presented in Eq.\ (\ref{mono4}) and Eq.\ (\ref{mono5}) it is straightforward to deduce one of our main results, a law to the QD distribution in arbitrary multi-partite pure states. For an arbitrary N-partite system, we can write
\begin{multline}
	\delta_{1|L_1}^{\leftarrow} + \delta_{2|L_2}^{\leftarrow} + \cdots + \delta_{N|L_N}^{\leftarrow}\\
	= \delta_{1|R_1}^{\leftarrow} + \delta_{2|R_2}^{\leftarrow} + \cdots + \delta_{N|R_N}^{\leftarrow}
	\label{eq:G_conservation_discord}
\end{multline}
where the notation $L_i$ means all the $(N-2)$ parts on left of the part $i$, that is $\{i-1,i-2,\dots, 1, N - 1, \dots, i+2\}$. For instance, $L_2$, means all the subsystems $\{1,N,N-1,\dots, 3\}$. Similarly, $R_i$ means the $N-2$ parts on right of the part $i$. 
We note that, given a system with $N$-parts, the sum of inaccessible information of each $i$-th part, given an observation on the rest, with exception of that immediately on right of $i$, is equal to the sum of inaccessible information of each $i$-th part, given an observation on the rest, with exception of that immediately on left of $i$. 
An intuitive illustrative scheme, is showed in Figure \ref{fig1}, where we show the idea exposed above for one part. There, on inset (a) we show the inaccessible information on part $1$, given an observation on the rest, with exception of that immediately on its left, \emph{i.e.}\ part $n$. On inset (b) we show again the inaccessible information on part $1$, but now given an observation on the rest,  with exception of that immediately on its right, \emph{i.e.}\ part $2$. The monogamous equality is reached doing the same procedure for all $n$ parts. These results generalizes the conservation law for QD considering multipartite systems.

\subsection{Conservation law with a measurement in one part}

As a final result, we present another simple {monogamy-like} conservation law that can be derived generalizing the of four-partite systems.
Consider the four equations:
\begin{eqnarray}
E_{bc|a} & = & \delta_{bc|d}^\leftarrow + S_{bc|d}, \nonumber \\
E_{cd|b} & = & \delta_{cd|a}^\leftarrow + S_{cd|a}, \nonumber \\
E_{da|c} & = & \delta_{da|b}^\leftarrow + S_{da|b}, \nonumber \\
E_{ab|d} & = & \delta_{ab|c}^\leftarrow + S_{ab|c}, 
\label{four21a}	
\end{eqnarray}
where the measurement in the discord is made only in one part.
Summing up Eqs.\ (\ref{four21a}), the conditional entropies cancel and we get	
\begin{multline}
	E_{bc|a} + E_{cd|b} + E_{da|c} + E_{ab|d} = \\
	\delta_{bc|d}^\leftarrow + \delta_{cd|a}^\leftarrow + \delta_{da|b}^\leftarrow + \delta_{ab|c}^\leftarrow.
	\label{claw4a}
\end{multline}
Therefore, Eq.\  (\ref{claw4a}) is a conservation law with the property that all measurements in the discords are made only in one subsystem.

The Eq.\ (\ref{claw4a}) can be easily generalized for an arbitrary number of parts. For that, let us consider a system of $N$ parts and label them with number from 1 to $N$, instead of letters.
We can write the following set of equations:
\begin{align}	\label{fundamentals_for_clawN}
	\hspace{1.5cm} E_{2:N-1|1} & =  \delta_{2:N-1|N}  & &+ S_{2:N-1|N}, \hspace{3cm} \nonumber\\
	E_{3:N|2} & = \delta_{3:N|1}  & &+  S_{3:N|1}, \nonumber\\
	E_{4:1|3} & = \delta_{4:1|2} & &+ S_{4:1|2}, \\
	\vdots \hspace{.3cm} & = \hspace{.3cm} \vdots & &+ \hspace{.3cm}\vdots \nonumber\\
	E_{1:N-2|N} & = \delta_{1:N-2|N-1} & &+ S_{1:N-2|N-1}, \nonumber
\end{align}
where the notation $X:Y$ means all the subsystems between numbers $X$ and $Y$ when $X<Y$. When $X>Y$, it meas all subsystems from $Y$ to $N$ and $1$ to $X$. Summing up Eqs. (\ref{fundamentals_for_clawN}), all the entropies cancel out and the result is the following conservation law between entanglement and discord:
\begin{multline}	\label{clawN}
	E_{2:N-1|1} + E_{3:N|2} + E_{4:1|3} + \cdots + E_{1:N-2|N} \\
		= \delta_{2:N-1|N} + \delta_{3:N|1} + \delta_{4:1|2} + \cdots + \delta_{1:N-2|N-1}.		
\end{multline}
This generalized conservation law shows a relation between EF and QD when just one part is measured, as we can note by the right side of Eq. (\ref{clawN}). {Although the bipartitions in Eq.\ (\ref{clawN}) overlaps, it also shows when the sum of quantum communications needed to form the correlations in the respective bipartitions are equal to the sum of locally inaccessible information.}

\section{Conclusion}
The way quantum correlations are distributed in a multipartite quantum systems is an aspect of great interest. It is well known that the distribution of correlations can not be made freely and understanding how this mechanism works has implications in the study of the monogamy of quantum correlations as well as in the understanding of protocols and other fundamental aspects of quantum information. Here we present a set of {monogamy-like} conservative laws that govern how the EF and the QD are distributed in multipartite systems. 
{These equalities links the constrains in the distributed entanglement with the distributed discord and vice-versa, showing that the monogamous properties of these two measures is deeply connected}.

We initially focus on four and five-partite systems and after we extend our results to multipartite systems. We show not only a general form of how the EF and the QD are distributed, but also a closed expression that rules how QD is distributed in multipartite systems.   These results elucidate important aspects in the distribution of quantum correlation in systems of many parts and may, in the near future, bring several implications and understandings to the quantum information theory. 

\bibliographystyle{apsrev4-1}
\bibliography{artigo}

\end{document}